\documentstyle[epsfig,12pt,a4p]{article}

\newcommand{\pipipi}{\mbox{$\pi^+\pi^-\pi^0$ }}
\newcommand{\etapipi}{\mbox{$\eta\pi^+\pi^-$ }}

\begin{document}
\begin{titlepage}
\def\footnoterule{\hrule width 1.0\columnwidth}
\begin{tabbing}
put this on the right hand corner using tabbing so it looks
 and neat and in \= \kill
\> {25 March 1998}
\end{tabbing}
\bigskip
\bigskip
\begin{center}{\Large  {\bf A study of
pseudoscalar states
produced centrally
in pp interactions at 450 GeV/c}
}\end{center}
\bigskip
\bigskip
\begin{center}{        The WA102 Collaboration
}\end{center}\bigskip
\begin{center}{
D.\thinspace Barberis$^{  5}$,
W.\thinspace Beusch$^{   5}$,
F.G.\thinspace Binon$^{   7}$,
A.M.\thinspace Blick$^{   6}$,
F.E.\thinspace Close$^{  4}$,
K.M.\thinspace Danielsen$^{ 12}$,
A.V.\thinspace Dolgopolov$^{  6}$,
S.V.\thinspace Donskov$^{  6}$,
B.C.\thinspace Earl$^{  4}$,
D.\thinspace Evans$^{  4}$,
B.R.\thinspace French$^{  5}$,
T.\thinspace Hino$^{ 13}$,
S.\thinspace Inaba$^{   9}$,
A.V.\thinspace Inyakin$^{  6}$,
T.\thinspace Ishida$^{   9}$,
A.\thinspace Jacholkowski$^{   5}$,
T.\thinspace Jacobsen$^{  12}$,
G.T\thinspace Jones$^{  4}$,
G.V.\thinspace Khaustov$^{  6}$,
T.\thinspace Kinashi$^{  14}$,
J.B.\thinspace Kinson$^{   4}$,
A.\thinspace Kirk$^{   4}$,
W.\thinspace Klempt$^{  5}$,
V.\thinspace Kolosov$^{  6}$,
A.A.\thinspace Kondashov$^{  6}$,
A.A.\thinspace Lednev$^{  6}$,
V.\thinspace Lenti$^{  5}$,
S.\thinspace Maljukov$^{   8}$,
P.\thinspace Martinengo$^{   5}$,
I.\thinspace Minashvili$^{   8}$,
K.\thinspace Myklebost$^{   3}$,
T.\thinspace Nakagawa$^{  13}$,
K.L.\thinspace Norman$^{   4}$,
J.M.\thinspace Olsen$^{   3}$,
J.P.\thinspace Peigneux$^{  1}$,
S.A.\thinspace Polovnikov$^{  6}$,
V.A.\thinspace Polyakov$^{  6}$,
V.\thinspace Romanovsky$^{   8}$,
H.\thinspace Rotscheidt$^{   5}$,
V.\thinspace Rumyantsev$^{   8}$,
N.\thinspace Russakovich$^{   8}$,
V.D.\thinspace Samoylenko$^{  6}$,
A.\thinspace Semenov$^{   8}$,
M.\thinspace Sen\'{e}$^{   5}$,
R.\thinspace Sen\'{e}$^{   5}$,
P.M.\thinspace Shagin$^{  6}$,
H.\thinspace Shimizu$^{ 14}$,
A.V.\thinspace Singovsky$^{ 1,6}$,
A.\thinspace Sobol$^{   6}$,
A.\thinspace Solovjev$^{   8}$,
M.\thinspace Stassinaki$^{   2}$,
J.P.\thinspace Stroot$^{  7}$,
V.P.\thinspace Sugonyaev$^{  6}$,
K.\thinspace Takamatsu$^{ 10}$,
G.\thinspace Tchlatchidze$^{   8}$,
T.\thinspace Tsuru$^{   9}$,
M.\thinspace Venables$^{  4}$,
O.\thinspace Villalobos Baillie$^{   4}$,
M.F.\thinspace Votruba$^{   4}$,
Y.\thinspace Yasu$^{   9}$.
}\end{center}

\begin{center}{\bf {{\bf Abstract}}}\end{center}

{
A study has been made of
pseudoscalar mesons produced centrally in pp interactions.
The results show that
the $\eta$ and $\eta^\prime$ appear to have a similar
production mechanism which differs from
that of the $\pi^0$.
The production properties of the $\eta$ and $\eta^\prime$ are not
consistent with what is expected from
double Pomeron exchange.
In addition the production mechanism for the $\eta$ and $\eta^\prime$
is such that the production cross section are greatest when
the azimuthal angle between the $p_T$ vectors of the two protons is
90~degrees.
}
\bigskip
\bigskip
\bigskip
\bigskip\begin{center}{{Submitted to Physics Letters}}
\end{center}
\bigskip
\bigskip
\begin{tabbing}
aba \=   \kill
$^1$ \> \small
LAPP-IN2P3, Annecy, France. \\
$^2$ \> \small
Athens University, Physics Department, Athens, Greece. \\
$^3$ \> \small
Bergen University, Bergen, Norway. \\
$^4$ \> \small
School of Physics and Astronomy, University of Birmingham, Birmingham, U.K. \\
$^5$ \> \small
CERN - European Organization for Nuclear Research, Geneva, Switzerland. \\
$^6$ \> \small
IHEP, Protvino, Russia. \\
$^7$ \> \small
IISN, Belgium. \\
$^8$ \> \small
JINR, Dubna, Russia. \\
$^9$ \> \small
High Energy Accelerator Research Organization (KEK), Tsukuba, Ibaraki 305,
Japan. \\
$^{10}$ \> \small
Faculty of Engineering, Miyazaki University, Miyazaki, Japan. \\
$^{11}$ \> \small
RCNP, Osaka University, Osaka, Japan. \\
$^{12}$ \> \small
Oslo University, Oslo, Norway. \\
$^{13}$ \> \small
Faculty of Science, Tohoku University, Aoba-ku, Sendai 980, Japan. \\
$^{14}$ \> \small
Faculty of Science, Yamagata University, Yamagata 990, Japan. \\
\end{tabbing}
\end{titlepage}
\setcounter{page}{2}
\bigskip
\par
The WA102 experiment is designed to study the properties of
centrally produced mesons.
Reactions of this type are expected to be
mediated by double exchange processes, with a mixture of Pomeron-Pomeron,
Reggeon-Pomeron and Reggeon-Reggeon exchange.
In order to try to understand the nature of the centrally produced mesons
it is interesting to try to determine which exchange process is producing
them. For instance,
the Pomeron
is thought to be a multi-gluonic object and
consequently it has been
anticipated that the production of
glue rich states may be especially favoured in
Double Pomeron Exchange (DPE).
It has been proposed
that the
$\eta$ and $\eta^\prime$
are strongly connected to glue~\cite{rosner}.
Further information on this subject could
come from studying if the $\eta$ or $\eta^\prime$ are produced by DPE.
\par
It has previously been reported that
in central production
$0^{-+}$ states
are suppressed relative to $1^{++}$
states~\cite{cenetapipi}.
In addition,
both the $\eta$ and $\eta^{\prime}$ signals are suppressed at small
four-momentum transfers where DPE is believed
to be dominant~\cite{cenetapipi,cen3pi}. Hence it could be conjectured
that $0^{-+}$ objects do not couple to DPE.
This hypothesis has been further tested by measuring the
cross section of the production of
the $\eta$ and $\eta^{\prime}$ as a function of
energy~\cite{new3pi,ckeiota}. The result shows that the
cross section for production of these resonances decreases with energy
which is inconsistent with production via DPE which is predicted
to give a constant cross section.
\par
This paper presents new results from the WA102 experiment on
the centrally produced $\pi^0$, $\eta$ and $\eta^\prime$ formed in the
reactions
\begin{equation}
pp \rightarrow p_{f} (\pi^0\rightarrow \gamma \gamma) p_{s}
\label{eq:a}
\end{equation}
\begin{equation}
pp \rightarrow p_{f} (\eta \rightarrow \gamma \gamma) p_{s}
\label{eq:b}
\end{equation}
\begin{equation}
pp \rightarrow p_{f} (\eta \rightarrow \pi^+\pi^-\pi^0, \pi^0 \rightarrow
\gamma \gamma) p_{s}
\label{eq:c}
\end{equation}
\begin{equation}
pp \rightarrow p_{f} (\eta^\prime \rightarrow \pi^+\pi^-\eta, \eta  \rightarrow
\gamma \gamma) p_{s}
\label{eq:d}
\end{equation}
at 450 GeV/c.
The subscripts f and s indicate the
fastest and slowest particles in the laboratory respectively.
The data come from experiment WA102
which has been performed using the CERN Omega Spectrometer.
The layout of the Omega Spectrometer used in this run is similar to that
described in ref.~\cite{wa9192} with the replacement of the
OLGA calorimeter by GAMS~4000~\cite{gams}.
\par
Reactions~(\ref{eq:a})-(\ref{eq:d})
have been isolated from the sample of events having the required number of
outgoing
charged tracks plus two $\gamma$s 
reconstructed in the electromagnetic
calorimeter\footnote{The showers associated with the impact of
the charged tracks on the calorimeter
have been removed from the event before the
requirement of only two $\gamma$s was made.}
by first imposing the following cuts on the components of the
missing momentum:
$|$missing~$P_{x}| <  17.0$ GeV/c,
$|$missing~$P_{y}| <  0.16$ GeV/c and
$|$missing~$P_{z}| <  0.12$ GeV/c,
where the x axis is along the beam
direction.
A correlation between
pulse-height and momentum
obtained from a system of
scintillation counters was used to ensure that the slow
particle was a proton.
Evidence of proton excitation can be observed in the
$p_{f} \pi^{0}$ mass spectrum (i.e. $\Delta^+$ or $N^{*+}$ production)
which has been removed by requiring
$M(p_{f} \pi^{0}) > 2.0 $ GeV.
\par
Fig.~\ref{fi:1}a) and b)  shows the two photon mass spectra
in the region of the $\pi^0$ and $\eta$ respectively for
reactions~\ref{eq:a} and~\ref{eq:b}.
The mass resolution is $\sigma= 14$~MeV for the $\pi^0$
and $\sigma= 33$~MeV for the $\eta$.
Fig.~\ref{fi:1}c) shows the \pipipi mass spectrum
in the region of $\eta$ for reaction~(\ref{eq:c})
where the $\pi^0$ is detected
decaying to two photons and kinematically fitted to give the $\pi^0$ mass.
The mass resolution is
$\sigma= 11$~MeV.
Fig.~\ref{fi:1}d) shows the \etapipi mass spectrum
in the region of $\eta^\prime$ for reaction~(\ref{eq:d})
where the $\eta$ is detected
decaying to two photons and kinematically fitted to give the $\eta$ mass.
The mass resolution is
$\sigma= 10$~MeV.
\par
A Monte Carlo simulation has been performed to calculate the
acceptance for each decay mode.
The acceptance corrected Feynman x ($x_F$) distributions are shown in
fig.~\ref{fi:2}a), b) and c) for the
$\pi^0$, $\eta$ and $\eta^\prime$ respectively for $x_F \geq 0.0$.
A cut of $0.0 \leq x_F \leq 0.1$ has been used
to select central events in a region where the acceptance is effectively flat
for all three mesons.
\par
After acceptance correction
the branching ratio of the $\eta$ to 2$\gamma$ and \pipipi has been calculated
to be
\begin{equation}
\frac{\eta \rightarrow \gamma \gamma}{\eta \rightarrow \pi^+\pi^-\pi^0} = 1.60
\pm 0.05 \pm 0.08
\end{equation}
which is consistent with the PDG~\cite{PDG96} value of $1.69\pm0.04$.
The systematic error is due mainly to the different trigger efficiencies
for the decay of central systems to all photon final states relative
to final states involving charged particles.
Since the background below the $\eta$ decay to \pipipi is smaller
than in the 2$\gamma$ decay mode, it is the data from the \pipipi decay
mode which will be used in the following analysis.
However, the 2$\gamma$ decay mode gives consistent results.
\par
It has recently been suggested~\cite{fec,jmf}
that the central production of pseudoscalars could be explained
if the production mechanism is through the fusion of two vectors.
In particular ref.~\cite{jmf} has suggested that
$\gamma \gamma$ fusion may be the dominant
mechanism for pseudoscalars in central production.
To test this hypothesis, and provide further information on their production
properties,
the cross sections for the production of pseudoscalar mesons have been
calculated
for the Feynman x interval
$0.0 \leq x_F \leq 0.1$
taking into account the unseen decay modes.
The resulting cross sections are
\begin{tabbing}
abhfgshdfghagawfe \= sigma mdsd \= =mm \=1234safwefwefe pm 23mm\= nbmmmyyyy
\kill
\> $\sigma(\pi^0)$ \>=  \>17609 $\pm$  210 $\pm$ 2400 \> nb \\
\> $\sigma(\eta)$ \>=  \>1295 $\pm$  16 $\pm$ 120 \> nb \\
\> $\sigma(\eta^\prime)$ \>=  \>588 $\pm$  18 $\pm$ 60 \> nb.
\end{tabbing}
The dominant contribution to the systematic error comes from the uncertainty
in the sensitivity of the experiment.
These cross sections are
several orders of magnitude larger than what would be expected
for $\gamma \gamma$ fusion which we have calculated, using the method
described in ref.~\cite{NACH},
to be less than
1~nb in the kinematic window considered.
\par
The relative cross sections can be measured more precisely
due to the cancellation of common systematic errors and are
\begin{center}
$\sigma(\pi^0):\sigma(\eta):\sigma(\eta^\prime)=1:0.073\pm 0.005: 0.033 \pm
0.002$.
\end{center}
These ratios are in agreement with
$\sigma(\pi^0):\sigma(\eta):\sigma(\eta^\prime)=1:0.080\pm 0.015: 0.028 \pm
0.009$
found previously by the
NA12/2 experiment studying the reaction $\pi^- N \rightarrow \pi^-N X^0$
at 300 GeV/c incident beam momentum~\cite{na120mp}.
It is interesting to note that
the NA12/2 experiment found that the $x_F$ dependence
of the
$\pi^0$, $\eta$ and $\eta^\prime$  was effectively flat in the range
$0.0 \leq x_F \leq 0.3$ in contrast to what is observed in this current
experiment (see fig.~\ref{fi:2}a), b) and c)).
\par
In previous analyses of other channels it has been observed that
when the centrally produced system has been analysed
as a function of the parameter $dP_T$, which is the difference
in the transverse momentum vectors of the two exchange particles~\cite{WADPT},
all the undisputed $q \overline q$ states are suppressed at small
$dP_T$.
Therefore, a study of the centrally produced pseudoscalar
states has been performed as a function of $dP_T$.
Fig.~\ref{fi:2}d), e) and f) shows the $dP_T$ spectra
for the $\pi^0$, $\eta$ and $\eta^\prime$ respectively.
The $dP_T$ spectra for the $\eta$ and $\eta^\prime$
are very similar; however, the $\pi^0$ has a different spectrum with more
events at small $dP_T$.
Table~\ref{ta:1} gives the fraction of each resonance in three bins of
$dP_T$.
As can be seen all the states are suppressed at small $dP_T$
consistent with what has been observed previously for all
other $q \overline q$ states~\cite{WADPT}.
This suppression of pseudoscalar states at small $dP_T$ is consistent
with their production through vector vector fusion~\cite{fec,jmf}.
\par
Fig.~\ref{fi:3}a), b) and c) shows the four momentum transfered from
one of the proton vertices
for the $\pi^0$, $\eta$ and $\eta^\prime$ respectively.
None of these distributions can be fitted with a single
exponential.
The $\eta$ and $\eta^\prime$ have a similar distribution while the
$\pi^0$ is different.
It should be noted that the $t$ spectra for the $\eta$ and $\eta^\prime$ are
not
consistent with
what is expected from DPE, namely an $e^{-bt}$ behaviour with
$b\approx 6-8 GeV^{-2}$~\cite{dpet}.
The t distributions have been fitted to the form
\begin{equation}
\frac{d\sigma}{dt} = \alpha e^{-b_1t} + \beta t^n e^{-b_2t}
\end{equation}
where values of n of 1 or 2 have been tried.
Fits with n~=~2 were found to give the best fit to the data.
The parameters resulting from the fit are given in table~\ref{ta:b}.
As can be seen the values of $b_1$ and $b_2$ are consistent in each case.
The main difference is that
the $\pi^0$ has a larger value of the ratio $\alpha/\beta$.
\par
The azimuthal angle ($\phi$) between the $p_T$
vectors of the two protons,
which is shown in
fig.~\ref{fi:3}d), e) and f), shows a very interesting effect. The $\eta$ and
$\eta^\prime$ are produced predominantly when the angle $\phi$
is 90 degrees.
\par
In summary,
it would appear that the $\eta$ and $\eta^\prime$ have a similar
production mechanism which differs from that of
the $\pi^0$.
The $t$ spectra for the $\eta$ and $\eta^\prime$ are not consistent with
what is expected from DPE and this fact
together with the previous observation that the
energy dependence of the cross section of the $\eta$ or $\eta^\prime$
is not constant~\cite{new3pi,ckeiota}
confirms that neither the $\eta$ nor the $\eta^\prime$
are produced dominantly by DPE.
In addition the production mechanism for the $\eta$ and $\eta^\prime$
is such that the production cross sections are greatest when
the azimuthal angle between the $p_T$ vectors of the two protons is
90~degrees.
\par
\begin{center}
{\bf Acknowledgements}
\end{center}
\par
This work is supported, in part, by grants from
the British Particle Physics and Astronomy Research Council,
the British Royal Society,
the Ministy of Education, Science, Sports and Culture of Japan
(grants no. 04044159 and 07044098)
and
the Russian Foundation for Basic Research (grant 96-15-96633).

\newpage
{ \large \bf Tables \rm}
\begin{table}[h]
\caption{Resonance production as a function of $dP_T$
expressed as a percentage of its total contribution.}
\label{ta:1}
\vspace{1in}
\begin{center}
\begin{tabular}{|c|c|c|c|} \hline
 & & &  \\
 &$dP_T$$\leq$0.2 GeV & 0.2$\leq$$dP_T$$\leq$0.5 GeV &$dP_T$$\geq$0.5 GeV\\
 & & & \\ \hline
 & & & \\
$\pi^0$  &13 $\pm$ 1 $\pm$ 2 & 45 $\pm$ 1 $\pm$ 2 &42 $\pm$ 1$\pm$ 4 \\
 & & & \\ \hline
 & & & \\
$\eta$  &6 $\pm$ 1$\pm$ 2 & 34 $\pm$ 1$\pm$2 &  60 $\pm$ 1 $\pm$ 3 \\
 & & & \\ \hline
 & & & \\
$\eta^\prime$  &3 $\pm$ 1 $\pm$ 2 & 33 $\pm$ 1 $\pm$ 2 &64 $\pm$ 1 $\pm$ 3 \\
 & & & \\ \hline
\end{tabular}
\end{center}
\end{table}
\newpage
\begin{table}[h]
\caption{Parameters from the fit to the t distribution}
\label{ta:b}
\vspace{1in}
\begin{center}
\begin{tabular}{|c|c|c|c|c|} \hline
 & & & & \\
 &$\alpha$ x$10^3$ & $b_1$ $GeV^{-2}$& $\beta$  x$10^3$& $b_2$ $GeV^{-2}$\\
 & & & &\\ \hline
 & & & & \\
$\pi^0$  & 128$\pm$ 9  &  14.9 $\pm$ 0.3 &  411$\pm$ 61 & 12.3 $\pm$ 0.4 \\
 & & & & \\ \hline
 & & & & \\
$\eta$  & 2.3 $\pm$ 0.3  & 15 $\pm$ 5  & 215 $\pm$12 & 11.6 $\pm$0.2  \\
 & & & & \\ \hline
 & & & & \\
$\eta^\prime$  & 0.3 $\pm$ 0.2  & 15 $\pm$ 5  & 108 $\pm$ 8 & 11.2 $\pm$0.2  \\
 & & & & \\ \hline
\end{tabular}
\end{center}
\end{table}
\newpage
\newpage
{ \large \bf Figures \rm}
\begin{figure}[h]
\caption{The effective mass spectra for the
reaction $pp \rightarrow p_fp_s X^0$ with $X^0$ decaying to
a), b) $\gamma \gamma$, c) $\pi^{+}\pi^{-}\pi^0$ and
d) $\eta\pi^+\pi^-$.}
\label{fi:1}
\end{figure}
\begin{figure}[h]
\caption{a), b) and c) The $x_F$ distribution for the $\pi^0$, $\eta$
and $\eta^\prime$ respectively.
d), e) and f) The $dP_T$ spectra for the $\pi^0$, $\eta$
and $\eta^\prime$ respectively.
}
\label{fi:2}
\end{figure}
\begin{figure}[h]
\caption{
The four momentum transfer squared ($|t|$) from one of the proton
vertices
for the a) $\pi^0$, b) $\eta$ and c) and $\eta^\prime$.
The azimuthal angle ($\phi$) between the two outgoing protons
for the d) $\pi^0$, e) $\eta$ and f) and $\eta^\prime$.
}
\label{fi:3}
\end{figure}
\newpage
\begin{center}
\epsfig{figure=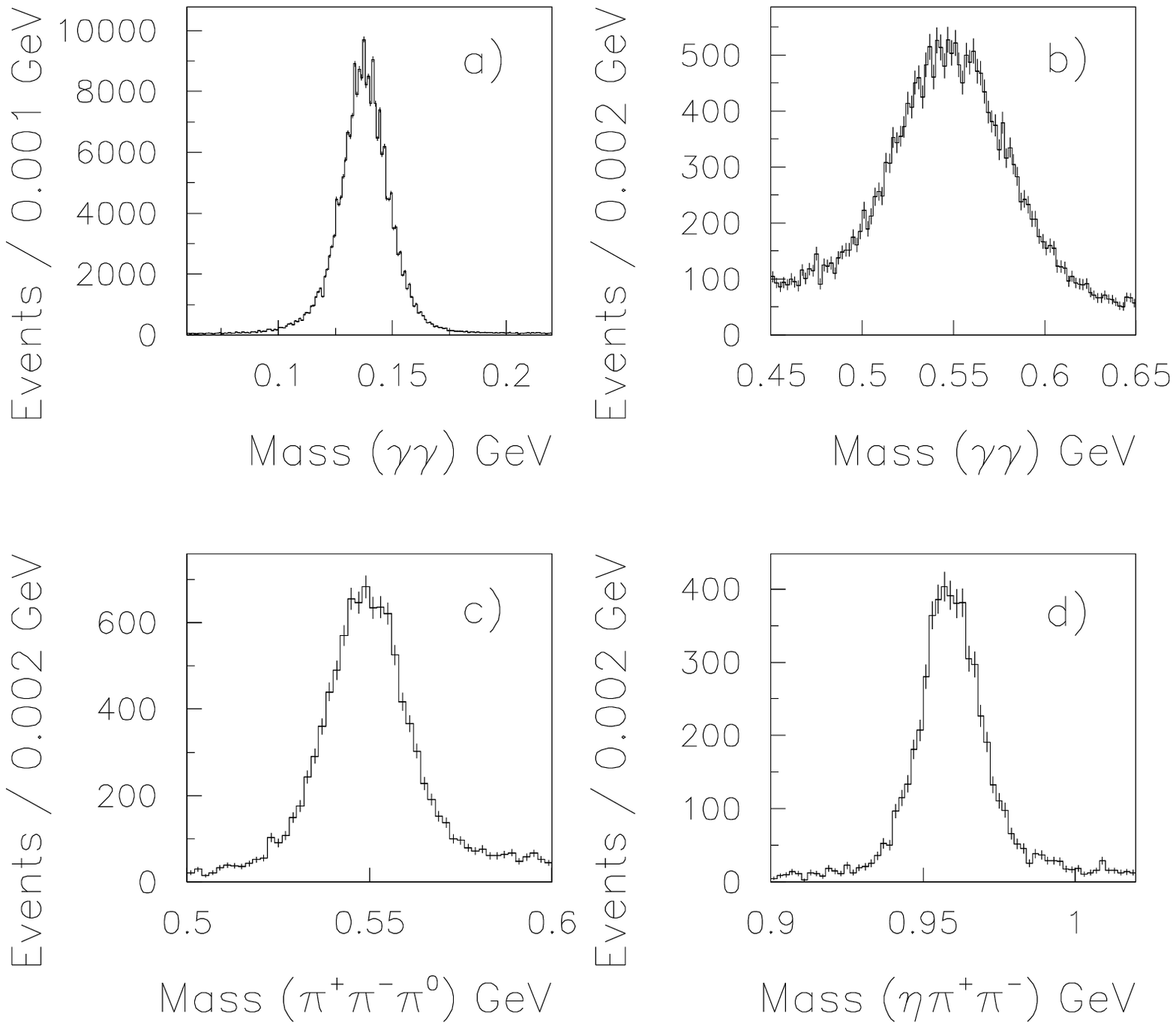,height=22cm,width=17cm}
\end{center}
\begin{center} {Figure 1} \end{center}
\newpage
\begin{center}
\epsfig{figure=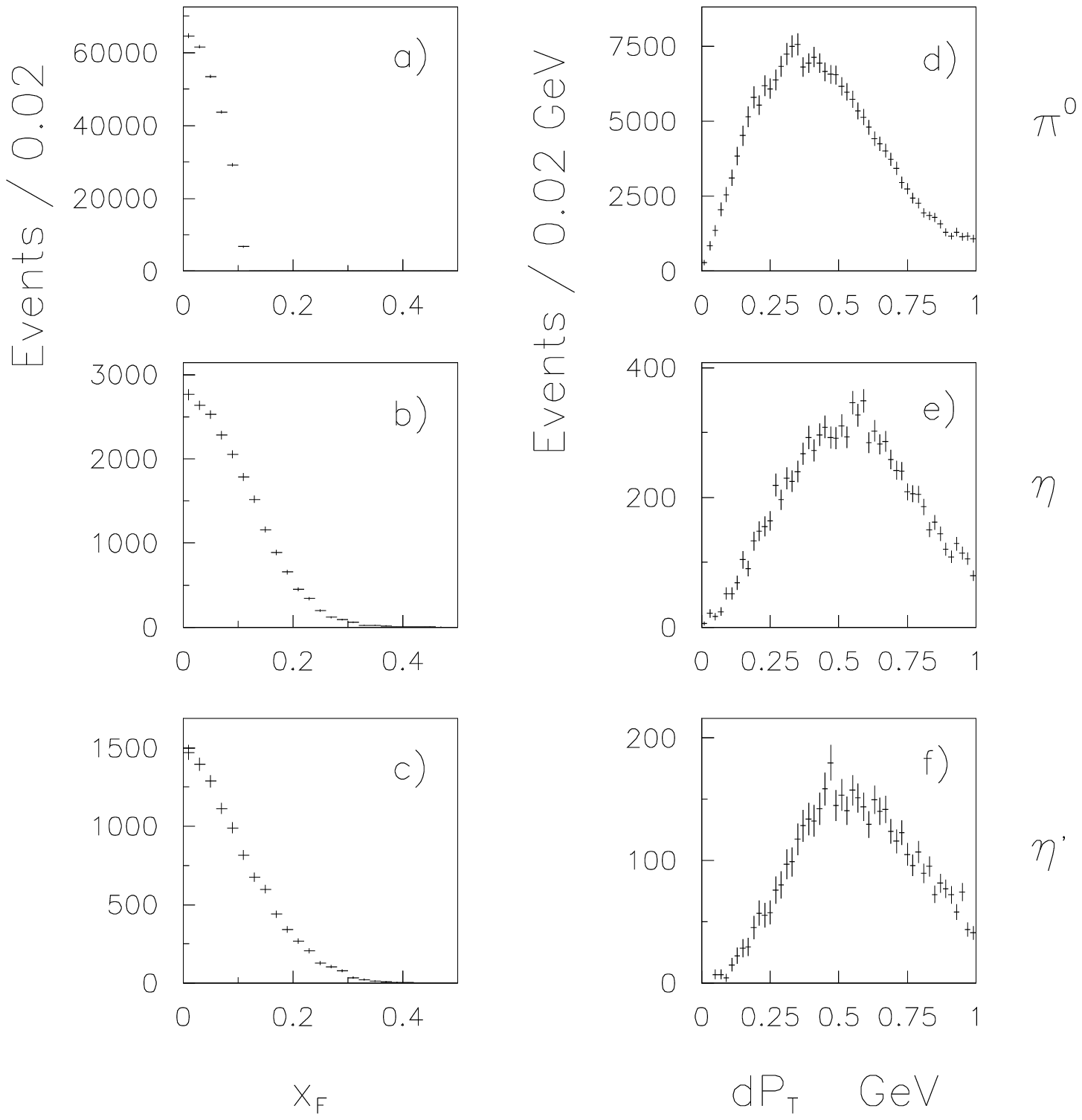,height=22cm,width=17cm}
\end{center}
\begin{center} {Figure 2} \end{center}
\newpage
\begin{center}
\epsfig{figure=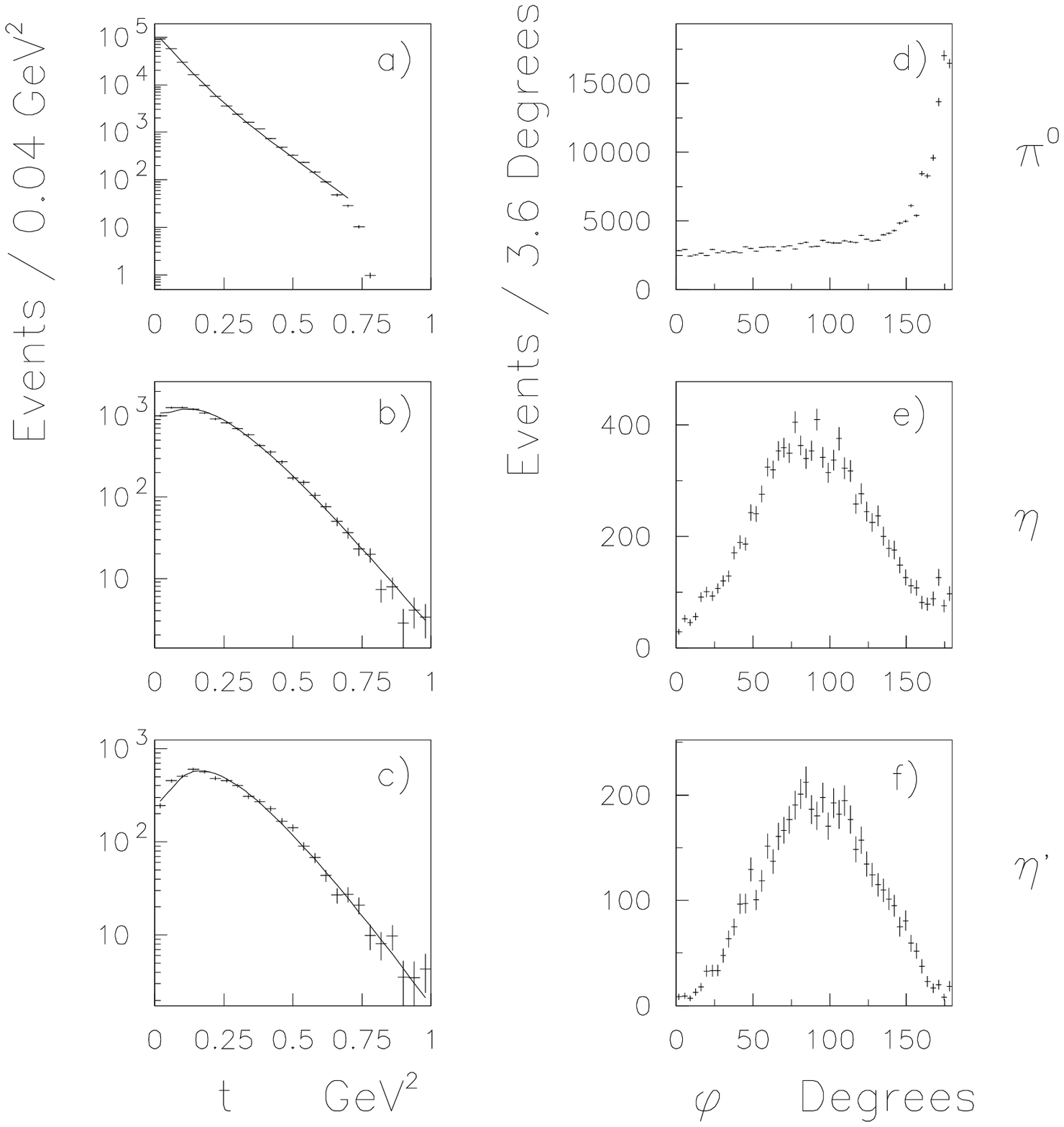,height=22cm,width=17cm}
\end{center}
\begin{center} {Figure 3} \end{center}
\end{document}